\documentclass[11pt,tightenlines,eqsecnum,floats,aps,prd,showpacs,amsmath,superscriptaddress,amssymb,nofootinbib,longbibliography]{revtex4-1}
\usepackage{graphicx, wrapfig}
\usepackage{amssymb}
\usepackage[usenames, dvipsnames]{color}
\usepackage{mathrsfs}
\usepackage{subfigure}
\usepackage{graphicx}
\usepackage{verbatim}
\usepackage{cancel}
\usepackage[normalem]{ulem}
\usepackage[hidelinks]{hyperref}
\hypersetup{
  colorlinks   = true, 
  urlcolor     = blue, 
  linkcolor    = blue, 
  citecolor   = blue 
}
\def\f{\frac}
\def\l{\left}
\def\r{\right}
\def\d{{\rm d}}
\def\pd{\partial}
\def\G{{\rm G}~}
\def\ie{i.e.~}
\def\vr{\overrightarrow{r}}
\def\vu{\overrightarrow{u}}
\def\vnabla{\overrightarrow{\nabla}}
\def\vx{\overrightarrow{x}}
\def\vv{\overrightarrow{v}}
\def\vR{\overrightarrow{R}}
\def\kpc{{\rm kpc}~}
\def\Msun{{\rm M_{sun}}}
\def\eV{{\rm eV}}
\def\pc{{\rm pc}}

\begin{document}
\title{Spherical collapse of fuzzy dark matter}
\author{V. Sreenath}\email{vsreenath@iucaa.in}

%
\affiliation{Inter-University Centre for Astronomy and Astrophysics, Post Bag 4, Ganeshkhind, Pune 411007, India.}
\begin{abstract}
It has been postulated that Fuzzy Dark Matter (FDM) could be a viable alternative 
to Cold Dark Matter (CDM).
FDM is comprised of ultralight bosons which exist as a Bose-Einstein condensate. 
Due to the very low mass of FDM, the de Broglie wavelength of these bosons are of the 
order of \kpc and the quantum effects manifest at those scales. 
Hence, unlike CDM, FDM experiences 
quantum pressure along with gravitational attraction.
In this work, we investigate the gravitational collapse of a spherically 
symmetric FDM overdensity. 
We assume a power law density profile for an overdense region of FDM and derive an expression for 
the temporal evolution of a spherical shell in the non-interacting limit and 
use it to derive an expression for average overdensity contained in the 
spherical shell in an Einstein--de Sitter universe. 
Further, we numerically extend the analysis to the case of interacting bosons.
Finally, we discuss the virialization of such an overdense region of non-interacting FDM and derive an expression 
for overdensity in the linear and the full theory. 
We compare our results with those obtained in the case of CDM and 
conclude with a discussion of the results.

\end{abstract}


\maketitle

\newpage
\section{Introduction} \label{sec:introduction}
Standard model of cosmology, namely the $\Lambda$CDM model has been a grand success \cite{Aghanim:2018eyx}. 
However this success also poses some serious questions. 
Of them a chief concern is regarding the nature of dark matter. 
Despite the success of Cold Dark Matter(CDM) at large scales, it has met with 
some problems at scales less than $10$ \kpc (for a recent review, see, for instance, \cite{Bullock:2017xww}). 
CDM predicts \cite{Navarro:1996gj} that the halos have a cusp in the density profile at its center. 
However, observations \cite{deBlok:2001hbg, deBlok:2005qh, Oh:2008ww} of low surface brightness galaxies and dwarf galaxies 
indicate that the density profile at the center of halos are shallower or in other 
words has a core (for a review, see, for instance, \cite{2010AdAst2010E...5D}).
Furthermore, simulations of CDM over predicts the number of dwarf galaxies in the 
local group by an order of magnitude \cite{Klypin:1999uc}. 
Of these two difficulties faced by CDM, it has been suggested that, the latter may be alleviated to an extent by 
taking in to account the effects due to baryons (for a recent work, see, for instance, \cite{2018arXiv180604143G}). 
However, the dust is yet to settle.
\par
In order to overcome the small scale issues of CDM, several alternatives to CDM has been suggested. 
One such alternative is Warm Dark Matter (WDM) (see, for instance, \cite{2001MPLA...16.1795J}). 
In this model, the dark matter particles possess a thermal velocity which causes them to free stream. 
This free streaming suppresses the formation of small scale structures thus solving the over abundance of 
dwarf galaxies and the core-cusp problem \cite{Colin:2007bk, SommerLarsen:1999jx}. 
However, the free streaming may also lead to certain imprints at large scale which can only be fixed by 
fine-tuning the parameters\cite{Narayanan:2000tp}. 
Another variant of CDM is the collisional dark matter (for a recent review, see, for instance, \cite{Salucci:2017cet}). 
It has been shown that the presence of collisions flattens the core and destroys the dwarf galaxies \cite{Spergel:1999mh}. 
However, an excess amount of collisions could also lead to the formation of singular core \cite{Yoshida:2000bx}. 
\par
Another proposal, which we will concern ourselves with in this article, is that the 
dark matter is composed 
of ultralight bosons \cite{PhysRevLett.85.1158}.
A popular candidate of such an ultralight bosonic dark matter, known commonly as Fuzzy Dark Matter (FDM), 
is axion of mass $m \sim 10^{-22} - 10^{-21}$ eV (for reviews on axion cosmology, see \cite{Sikivie:2006ni, Marsh:2015xka}).
All large scale properties of FDM is similar to that of CDM. 
However, at small scales quantum properties of FDM affects the formation of structure. 
Due to the small mass of FDM, the de Broglie wavelength is of order of \kpc. 
The de Broglie wavelength manifests itself as a Jeans length below which 
the quantum pressure due to the uncertainty principle acts against gravity.
Thus, below the de Broglie wavelength, the pressure suppresses the formation of 
structure and flattens the density profile\cite{PhysRevLett.64.1084, Sin:1992bg, Sahni:1999qe, PhysRevLett.85.1158}.  
The implications of FDM model to structure formation has been investigated (see, for instance, \cite{Marsh:2013ywa, Hui:2016ltb}). 
Most of the current searches of dark matter are not designed to detect FDM and hence the negative results do not 
constrain it. However, experiments have been proposed which are likely to detect FDM 
\cite{Arvanitaki:2009fg, Kim:2015yna, Graham:2013gfa, Stadnik:2014tta, Stadnik:2015kia, Abel:2017rtm, Hees:2018fpg}. 
\par
In the FDM model, the ultralight bosons form a Bose-Einstein Condensate (BEC) at a very early time. 
In a BEC all the dark matter particles occupy the ground state and hence is described by a 
coherent wave function. 
The evolution of such a system is described by the Schr{\"o}dinger equation together with 
an equation governing gravity. 
The exact dynamics of structure formation can only be explored using numerical simulations \cite{Woo:2008nn, Schive:2014dra, Nori:2018hud, Veltmaat:2018dfz}. 
High-resolution numerical simulations \cite{Schive:2014dra} show that the halo centers have a solitonic core with 
the outer profile similar to NFW \cite{Navarro:1996gj}. 
Even though numerical simulations are required to have a complete understanding of the 
structure formation, analytical approximations often give useful insights. 
In this spirit, there has been analytical investigations of the steady state of a spherically symmetric 
Newtonian self-gravitating BEC \cite{PhysRev.187.1767, Boehmer:2007um, 2011PhRvD..84d3531C, 2011PhRvD..84d3532C, Chavanis:2011cz, Harko:2014vya, Chavanis:2016dab}. 
In these studies, the nature of the virialized halo has been investigated, either assuming that the system 
is in hydrostatic equilibrium or by using virial theorem.  
For some other approaches to the study of collapse of axionic scalar field or formation of structure from 
axions, see, for instance, \cite{deFreitas:2015dwa, Eby:2016cnq, Schiappacasse:2017ham, 1985MNRAS.215..575K, 1994PAN....57..485S, Sakharov:1996xg, Khlopov:1998uj}.
In this work, following the footsteps of an earlier work on CDM \cite{1972ApJ...176....1G}, 
we would like to study the gravitational collapse of a spherically overdense region of FDM. 
Firstly, assuming a spherically symmetric power law density profile, we will analytically investigate the 
time evolution of a spherical shell of FDM comprising of non-interacting bosons. 
We will use the analytical solutions for radius of the shell to arrive at an expression 
for overdensity and study it in the linear regime. 
Secondly, we will extend the analysis numerically to the case of interacting bosons, and study the 
evolution of a spherical shell and its dependence on the strength of interactions. 
Finally, we will use virial theorem for non-interacting FDM to compute the critical 
overdensity at which the spherically overdense region would collapse in to a halo.
\par
The rest of the paper is organized as follows. In the next section we will describe 
the Gross-Pitaevskii-Poisson system that governs a BEC evolving under the effect 
of gravity. 
We describe how such a system can be expressed as hydrodynamic equations, namely the 
continuity, Euler and Poisson equations. 
From the hydrodynamic equations, we derive the equation of motion of a spherical shell containing an 
overdense region of FDM. 
In section III, assuming a density profile for the overdense region, we analytically and numerically 
solve for the equation of motion of the 
spherical shell to calculate the evolution of the shell with time 
and use the analytical expression to 
compute the expression for averaged overdensity contained in the spherical shell as a 
function of time. 
We further compute the evolution of the shell, numerically, for the case of interacting bosons.  
In section IV, we use virial theorem to compute the critical overdensity at which the overdense region 
will collapse in to a halo.
We conclude the paper in section V with a discussion of the results. 
\par
We shall work with the units where $c\,=\,1$.

\section{Gross-Pitaevskii-Poisson system and equation for spherical collapse}
FDM is a BEC evolving under the effect of gravity. 
Since, general relativistic effects are not significant at the scale of a halo we 
can use Newtonian gravity. 
The state of such a system is described by the condensate wave function, $\psi(t,\,\vr)$, 
governed by the Gross-Pitaevskii-Poisson (GPP) equations (see, for instance, \cite{Dalfovo:1999zz, 2012A&A...537A.127C}),
\begin{subequations}
 \begin{eqnarray}
 i\hbar\f{\pd \psi(t,\,\vr)}{\pd t} &=& - \f{\hbar^2}{2\,m}\nabla^2\psi(t,\,\vr)\, +\, m\, \Phi(t,\,\vr)\,\psi(t,\,\vr)\, +\, \f{4\,\pi\,a_s\,\hbar^2}{m^2} |\psi(t,\,\vr)|^2\psi(t,\,\vr)\label{eqn:GP}\\
 \nabla^2 \Phi(t,\,\vr)\, &=&\, 4\, \pi\, \G\,|\psi(t,\,\vr)|^2,\label{eqn:Poisson1}
\end{eqnarray}\label{eqn:GPP}
\end{subequations}
where $m$ is the mass of the boson, $\rho(t,\,\vr) = |\psi(t,\,\vr)|^2$ is the mass density 
and $\Phi(t,\,\vr)$ is the gravitational potential.
The self-interaction of bosons is described by the s-wave scattering length, $a_s$, 
which can be positive (repulsive), negative (attractive) or zero (non-interacting). 
For brevity, whenever possible, we will not explicitly write the coordinate dependence of quantities.
\subsection{Hydrodynamic equations}
It is often convenient to express the GPP equations, describing the FDM halo, 
in terms of fluid variables, namely density and velocity \cite{2012A&A...537A.127C}. 
This can be achieved by performing a Madelung transformation \cite{Madelung1927}, 
\begin{equation}
 \psi(t,\,\vr) = \sqrt{\rho(t,\,\vr)}\exp(i\,S(t,\,\vr)/\hbar).\label{eqn:Madelung}
\end{equation}
Upon substituting Eq. (\ref{eqn:Madelung}) in Eq. (\ref{eqn:GP}), defining 
\begin{equation}
 \vu(t,\,\vr) \equiv \f{ \vnabla S(t,\,\vr)}{m},\label{eqn:u}
\end{equation}
equating real and imaginary parts of Eq. (\ref{eqn:GP}) and using the identity
\begin{equation*}
 (\vu \cdot \vnabla) \vu \, =\, \vnabla(u^2/2)\, -\, \vu \times(\vnabla \times \vu) = \, \vnabla(u^2/2),
\end{equation*}
we obtain
\begin{eqnarray}
\f{\pd \rho}{\pd t}  + \vnabla\cdot (\rho \vu) &=& 0, \label{eqn:continuity}\\
\f{\pd \vu}{\pd t} + ( \vu \cdot \vnabla)\vu &=& -\f{\vnabla P}{\rho} - \vnabla \Phi - \f{\vnabla Q}{m},\label{eqn:Euler}\\
\nabla^2\Phi &=& 4\,\pi\,\G\,\rho,\label{eqn:Poisson2}
\end{eqnarray}
which are, respectively, the continuity, Euler and Poisson equations of a fluid with density $\rho$ and velocity $\vu$. 
Equation (\ref{eqn:u}) implies that the velocity, $\vu$, of the fluid is irrotational. 
In the Euler equation, we have denoted the quantum potential by
\begin{equation}
 Q(t,\,\vr) = -\f{\hbar^2}{2\, m}\, \f{\nabla^2 \sqrt{\rho}}{\sqrt{\rho}} 
 = -\f{\hbar^2}{4\, m} \l[ \f{\nabla^2\rho}{\rho}\, -\, \f{1}{2}\f{(\nabla\rho)^2}{\rho^2} \r],
\end{equation}
and the pressure, arising due to self interactions, by
\begin{equation}
 P(t,\,\vr) = \f{2\,\pi\,a_s\,\hbar^2}{m^3}\,\rho^2.
\end{equation}
Note that the above equation describes an equation of state of a polytrope of index one.
\par
In an expanding universe, $\vr(t) = a(t) \vx$.
Using the relation,
\begin{equation}
 \l.\f{\pd}{\pd t}\r|_{\vr}\, =\, \l.\f{\pd}{\pd t}\r|_{\vx}\, -\, H \vx\cdot \vnabla,
\end{equation}
where $H(t) = \dot a(t)/a(t)$ is the Hubble parameter, the fluid equations can be written as
\begin{eqnarray}
 &&\f{\pd \rho}{\pd t}\, -\, H\,(\vx \cdot \vnabla)\,\rho\, +\, \f{\vnabla\cdot(\rho\,\vu)}{a}\, =\, 0,\\
 &&\f{\pd \vu}{\pd t}\, -\, H\,(\vx\cdot\vnabla)\vu\, + \f{( \vu \cdot \vnabla)\vu}{a}\, =\, -\f{\vnabla P}{a\, \rho} - \f{\vnabla \Phi}{a} - \f{\vnabla Q}{a\,m},\\
 && \nabla^2\Phi\, =\, 4\,\pi\,\G\,a^2\,\rho,
\end{eqnarray}
 where $\vnabla$ is now with respect to $\vx$,
 \par
 Let us now separate the density of the condensate in to a background part and a perturbation on top of it, \ie $\rho = \rho_b( 1 + \delta)$, where $\delta = \delta\rho/\rho_b$. 
 A similar split can be made to the velocity of the fluid element in to the Hubble flow and peculiar velocity, \ie $\vu\,=\,H\,\vr\,+\,\vv$, respectively.
 Using these definitions, one could write the perturbed part of the fluid equations as
 \begin{eqnarray}
  &&\f{\pd \delta}{\pd t}\, +\, \f{\vnabla}{a}\cdot \l[ \vv\,(1\, +\,\delta)\r]\, =\, 0,\label{eqn:continuity-perturbed}\\
  &&\f{\pd \vv}{\pd t}\, +\, H\,\vv\, +\, \f{(\vv\cdot\vnabla)\vv}{a}\, =\, -\f{4\,\pi\,a_s\,\hbar^2}{a\,m^3}\vnabla\rho\, 
  -\, \f{\vnabla\Phi_p}{a}\, +\, \f{\hbar^2}{4\,m^2\,a^3}\vnabla \l[ \f{\nabla^2\rho}{\rho}\, 
  -\, \f{1}{2}\f{(\nabla\rho)^2}{\rho^2} \r],\label{eqn:Euler-perturbed}\\
  &&\nabla^2\Phi_p\, =\, 4\,\pi\,\G\,a^2\,\rho_b\delta,
 \end{eqnarray}
In the above, we have also divided the gravitational potential, $\Phi\, =\, \Phi_b\, +\, \Phi_p$, in to background and perturbation parts respectively. 
It can be shown that $\Phi_b\, =\, -\ddot a\, r^2/ (2\, a)$. 
In writing the perturbed part of Euler equation, we have retained the full density, 
$\rho$, on the right-hand side for later convenience.
\subsection{Equation for a collapsing spherical shell}
Consider a spherically overdense distribution of FDM. 
We are interested in understanding the evolution of such an overdense region 
with time. 
Consider a spherical shell of radius $R(t) = a(t) X(t)$, enclosing certain mass, 
centered in the overdense region. 
A fluid element on that shell would have a velocity, $\vu\, =\, H\,\vR\, +\, \vv$, 
where the velocity of the fluid element is in radial direction. 
The acceleration of that fluid element can be computed as
\begin{equation}
 \f{\d^2 \vR}{\d t^2}\, =\, \f{\d \vu}{\d t}\,
 =\, \dot H\,\vR\, +\, H\,(H\,\vR\, +\, \vv)\, +\, \f{\pd \vv}{\pd t}\, +\, \f{(\vv\cdot\vnabla)}{a}\vv .
\end{equation}
Upon using the Euler equation, Eq. (\ref{eqn:Euler-perturbed}), and the fact that 
$\vnabla \Phi_b\, =\, -\ddot a\, \vR/a^2$, we obtain
\begin{equation}
 \f{\d^2 \vR}{\d t^2}\, =\,-\f{\vnabla\Phi_b}{a}\, -\f{4\,\pi\,a_s\,\hbar^2}{a\,m^3}\vnabla\rho\, 
 -\, \f{\vnabla\Phi_p}{a}\, +\, \f{\hbar^2}{4\,m^2\,a^3}\vnabla \l[ \f{\nabla^2\rho}{\rho}\, 
 -\, \f{1}{2}\f{(\nabla\rho)^2}{\rho^2} \r].
\end{equation}

Combining the background and perturbed parts of the gravitational potential,
one can write the equation of motion of the spherical shell as
\begin{equation}
 \f{\d^2 \vR}{\d t^2}\, =\, -\f{4\,\pi\,a_s\,\hbar^2}{m^3}\vnabla\rho\, 
 -\, \vnabla\Phi\, +\, \f{\hbar^2}{4\,m^2}\vnabla \l[ \f{\nabla^2\rho}{\rho}\, 
 -\, \f{1}{2}\f{(\nabla\rho)^2}{\rho^2} \r],\label{eqn:R-general}
\end{equation}
where the spatial derivatives are now with respect to $r$ and are evaluated on the shell, $r = R(t)$. 
Thus, the motion of the spherical shell is governed by a force arising due to the bosonic interactions which could be attractive ($a_s<0$) or repulsive ($a_s>0$),  gravitational attraction and a quantum repulsive force. 
It can be seen that, in the limit $\hbar/m \rightarrow 0$, we reproduce the equation for spherical 
collapse of CDM. 
In order to make further progress, we need to assume a density profile and study the evolution of the above equation.
\section{Spherical collapse of FDM with a power law density profile}
For the  overdense region, we consider a power law density profile of the form
\begin{equation}
 \rho(t,r)\, =\, \f{3\, -\, \gamma}{4\pi}\f{M}{L(t)^3}\l(\f{r}{L(t)}\r)^{-\gamma}\label{eqn:pl-profile},
\end{equation}
where the normalization factors has been chosen in such a way that, 
$L(t)$ is the radius of the shell which encloses a mass $M$ and $\gamma$ 
is a positive number less than $3$ (by demanding that density should be positive). 
Assuming that the FDM overdense region maintains such a density profile throughout the evolution, 
one can derive the equation of motion for the spherical shell.
For a shell of radius $L(t)$, containing mass $M$, the equation of motion can be written as 
\begin{equation}
 \f{\d^2 L}{\d t^2}\, =\, \gamma\,(3\,-\,\gamma)\f{\,a_s\,\hbar^2}{m^3}\f{M}{L^4}\, -\f{\G\,M}{L^2}\, + \gamma\,(2\,-\,\gamma)\f{\hbar^2}{4\,m^2\,L^3}. \label{eqn:eom}
\end{equation}
As explained before, the evolution of the shell is governed by three forces, namely, 
(i) the repulsive ($a_s>0$) or attractive ($a_s<0$) force due to bosonic self-interaction, 
(ii) attractive gravitational force and (iii) repulsive quantum force. 
Note that, for the power law profile, in order for the quantum force to be positive and non-vanishing, 
one requires $\gamma < 2$. 
\subsection{Non-interacting bosons}
We will investigate the case of non-interacting bosons in this subsection and consider the 
effects of interaction in the next.
\subsubsection{Analytical solution}\label{subsubsec:analytic}
In the absence of interactions ($a_s\,=\,0$), the equation of motion of the 
spherical shell can be written as,
\begin{equation}
m\,\f{\d^2 L}{\d t^2}\, =\, -\f{k}{L^2}\, +\, \f{l^2}{m\,L^3},\label{eqn:kepler}
\end{equation}
where $k\,=\,\G\,M\,m$ and $l^2\,=\,(2\,\gamma\, -\, \gamma^2)\,\hbar^2/4$. 
This equation is mathematically, though not physically, similar to the equation governing 
the reduced mass in a two-body Kepler problem (see, for instance, \cite{goldstein}). 
Hence, we will draw insights from the solution of Kepler problem to solve Eq. (\ref{eqn:kepler}).
\par
Initially, let the overdense shell be expanding along with the Hubble flow. 
The shell will eventually turn around if the initial value of the first integral 
of motion of the shell is negative, 
\ie if,
\begin{equation}
 E\, =\, \f{1}{2}\,m\,\l(\f{\d L}{\d t}\r)^2\, +\, \f{l^2}{2\,m\,L^2}\, -\, \f{k}{L}\, <\, 0. \label{eqn:E}
\end{equation}
In such a case, it can be shown that 
the solution to Eq. (\ref{eqn:kepler}) 
can be expressed as follows
\begin{subequations}
 \begin{eqnarray}
 L\, &=& A\l( 1\, -\, e\,\cos\vartheta \r),\label{eqn:sol-l}\\
 t\, &=&\, \l(\f{m\,A^3}{k}\r)^{1/2}\l( \vartheta\, -\, e\,\sin\vartheta \r),\label{eqn:sol-t}
\end{eqnarray}\label{eqn:sol-Lt}
\end{subequations}
where, $A\, = \, -k/(2\,E)$ and the expression for $e$ is
\begin{equation}
 e\, =\, \sqrt{1\, +\, \f{2\,E\,l^2}{m\,k^2}}\, =\, \sqrt{1\, +\, \f{E\, \hbar^2}{\G^2\,M^2\,m^3}\,\f{(2\,\gamma\, -\gamma^2)}{2}}.\label{eqn:e}
\end{equation}
Note that, since $E<0$, the value of $e <1$. 
The Eqs. (\ref{eqn:sol-Lt}) can be combined to obtain the behaviour of $L$ as a function of $t$. 
\par
Let us now try to understand the behaviour of the solution. 
When $\vartheta\,=\, 0$, $L(0)\, =\, L_{min}\,=\, A(1\,-\,e)$ and 
when $\vartheta\,=\, \pi$, $L(\pi)\, =\, L_{max}\,=\, A(1\,+\,e)$. 
Since $e<1$, the radius of the shell is thus bounded from below and 
hence will oscillate between the two extremum values. 
\begin{figure}
 \includegraphics[width = 12cm]{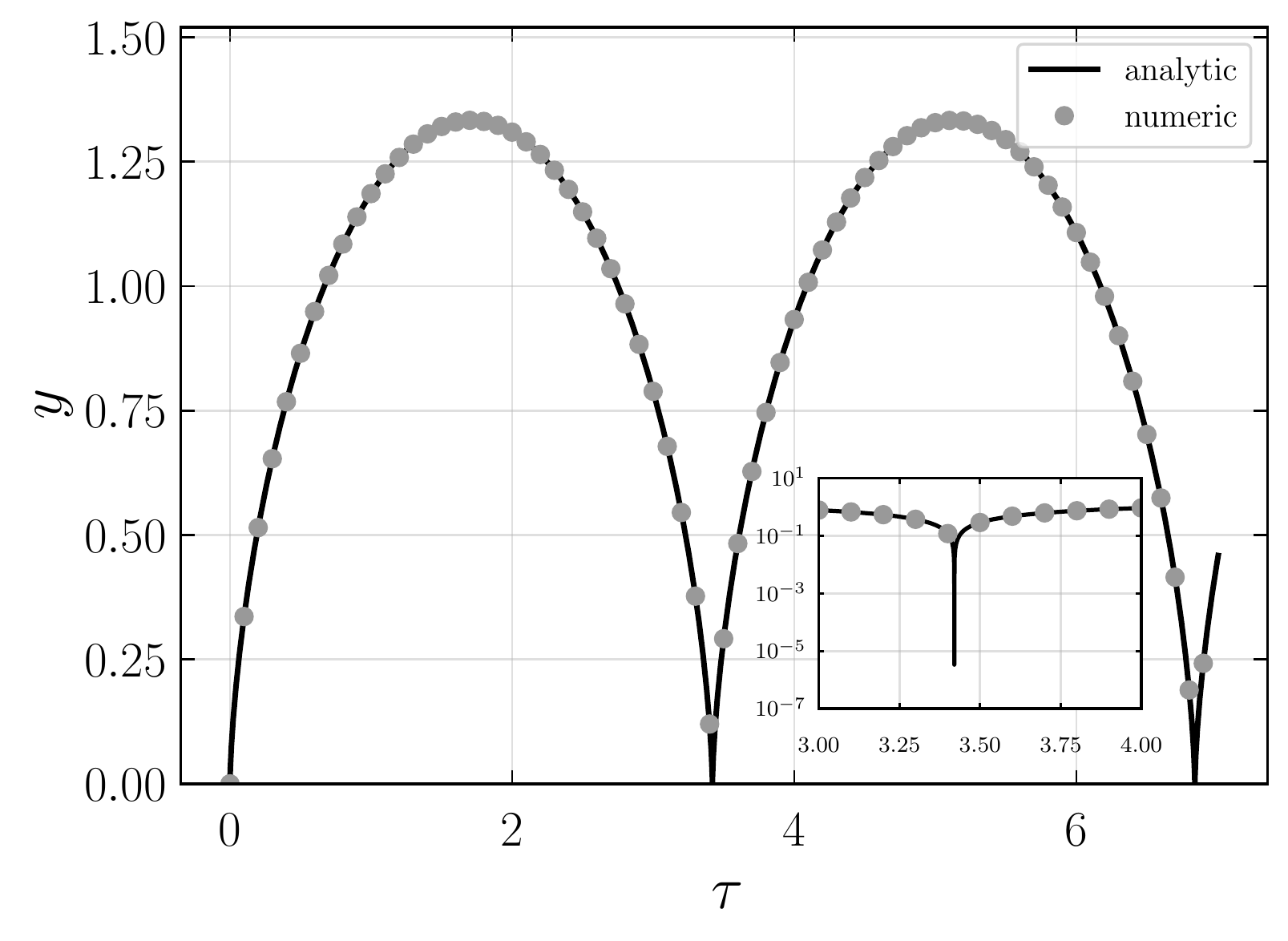}
 \caption{\label{fig:1}Comparison of analytical and numerical results for the 
 evolution of $y$ as a function of $\tau$. 
 As expected, the numerical and analytical results match very well. 
 For a shell containing a mass $M\,=\,9.1\times 10^7\,\Msun$ and for bosonic mass $m\,=\,8.1\times 10^{-23}\,\eV$, 
 this evolution corresponds to a shell oscillating between 
 a minimum radius of $L_{\rm min}\, 
=\, 3.57\times10^{-8}\,\pc$ 
and a maximum radius of $L_{\rm max}\, =\, 1.9\, \kpc$. 
}
\end{figure}
\subsubsection{Numerical solution}
Let us now solve Eq. (\ref{eqn:kepler}) numerically and compare it with the 
analytical solutions Eqs. (\ref{eqn:sol-Lt}). 
For numerical simulations, it is convenient to rewrite Eq. (\ref{eqn:kepler}) in terms of dimensionless 
variables as,
\begin{equation}
 \f{\d^2 y}{\d\tau^2}\, =\, \f{2\,\gamma\, -\, \gamma^2}{4\,y^3}\, -\, \f{1}{y^2} \label{eqn:y}
\end{equation}
where, we have defined $y\, =\, L/L_Q$ and $\tau\, =\, t/t_Q$, with 
$L_Q = \hbar^2/(G\,M\,m^2)$ and $t_Q\, =\, \sqrt{L_Q^3/(G\,M)}$. 
In terms of $y$ and $\tau$, we only need to solve the differential 
equation for a given initial condition once. The solution for 
the shell with radius $L(t)$ containing any mass $M$ 
can be then obtained from solution for $y(\tau)$ by scaling it with corresponding
$L_Q$ and $t_Q$.
\par
In order to solve for $y(\tau)$, we assume that, in the beginning, the shell 
containing an average overdensity of $\bar\delta_i\,=\, 10^{-5}$, 
is expanding according to the Hubble flow. 
Furthermore, we assume that the universe is Einstein--de Sitter(EdS); \ie 
the scale factor scales with time as $a\propto t^{2/3}$. 
Since we are interested in bound solutions, we assume that the first integral of 
motion of Eq. (\ref{eqn:y}) is less than zero. 
Using these three conditions, we can fix the initial conditions for Eq. (\ref{eqn:y}). 
For $\gamma\,=\,10^{-10}$, we have numerically solved for $y(\tau)$ using Mathematica \cite{Mathematica} 
and compared with the analytical solutions expressed in terms of $y(\tau)$. 
As can be seen from Fig. \ref{fig:1}, as expected, both the solutions match very well. 
In order to make sense of the numbers involved, let us assume that the shell contains a 
mass of $M\, =\, 9.1\times 10^7\, \Msun$ and that the mass of boson is 
$m\, =\, 8.1\times10^{-23}\, \eV$ (values arrived at for Fornax dwarf spheroidal in \cite{Schive:2014dra}). 
Such a shell, for the initial condition we have considered, would oscillate between 
a minimum radius of $L_{\rm min}\, 
=\, 3.57\times10^{-8}\,\pc$ 
and a maximum radius of $L_{\rm max}\, =\, 1.9\, \kpc$. 
We have considered an initial state for which $y_i\,=\, 10^{-5}$ and $\bar\delta_i\,=\, 10^{-5}$.
For this state, it is interesting to note that $1\,-\,e\, =\, {\cal O}(10^{-11})$, 
hence such a vast difference in $L_{\rm min}$ and $L_{\rm max}$.
\subsubsection{Overdensity}\label{subsubsec:overdensity}
Let us assume that the background spacetime is EdS.
The average density contained in a spherical shell of radius $L$ containing mass $M$ is given by $\bar\rho\, =\, M/(4\,\pi\,L^3/3)$. 
In an EdS universe, the background density is given by $\rho_b\, =\, 1/(6\,\pi\,\G\,t^2)$. 
Hence, the average overdensity inside the spherical shell is
\begin{equation}
 1\, +\, \bar\delta\, =\, \f{\bar\rho}{\rho_b}\, =\, \f{9}{2}\f{\G\,M\,t^2}{L^3}.
\end{equation}
Substituting Eqs. (\ref{eqn:sol-Lt}) for $L$ and $t$, we obtain
\begin{equation}
 1\, +\, \bar\delta\, =\, \f{9}{2}\,\f{(\vartheta\, -\, e\,\sin\vartheta)^2}{(1\, -\, e\,\cos\vartheta)^3}.\label{eqn:overdensity}
\end{equation}
The above expression for average overdensity within the shell has the following properties: 
(i) since $e<1$, the overdensity does not diverge as $\vartheta \rightarrow 2\pi$, 
(ii) the averaged overdensity is fluctuating and increasing with time,
(iii) in the limit $\hbar/m \rightarrow 0$, $e\rightarrow 1$, Eq. (\ref{eqn:overdensity}) reproduces the CDM expression 
(see, for instance, \cite{paddylss1993}) for 
averaged overdensity.

\par
Let us now turn our attention to the behaviour of $\bar\delta$ in the 
small $\vartheta$ limit. 
Upon Taylor expanding the expression for $\bar\delta(\vartheta)$, Eq. (\ref{eqn:overdensity}), about 
$\vartheta\, \simeq\, 0$, we obtain 
\begin{equation}
 1\, +\, \bar\delta\, \simeq\, \f{9}{2}\l(\f{\vartheta^2}{1\,-\,e}\r)\, 
 -\, \f{21\,e}{4}\,\l( \f{\vartheta^2}{1-e} \r)^2\, 
 +\, ...\, .
\end{equation}
The above expansion for $\bar\delta$ would be valid only if $\vartheta^2\, <<\, 1\,-\,e$. 
However, in this limit, the above expression imply that $\bar\delta\, \simeq -1$ which indicate 
an underdensity. 
If as we saw in the last section, $1\,-\,e$ is very small, then one could first take the limit 
of $e\rightarrow 1$ and then the limit $\vartheta\rightarrow 0$.
%
%
Upon taking the limit in this order, of Eqs. (\ref{eqn:sol-t}) and (\ref{eqn:overdensity}), we obtain,
\begin{equation}
 \bar\delta\, \simeq\, \f{3\,\vartheta^2}{20}\, \propto a,\label{eqn:lineardelta}
\end{equation}
which is similar to that in CDM. 
The above discussion seem to indicate that, in this model, for an overdense region, 
a sensible small $\vartheta$ limit exists only if 
the 
limit $e\rightarrow 1$ can be taken before the $\vartheta\rightarrow 0$ limit.

\begin{figure}
 \includegraphics[width = 12cm]{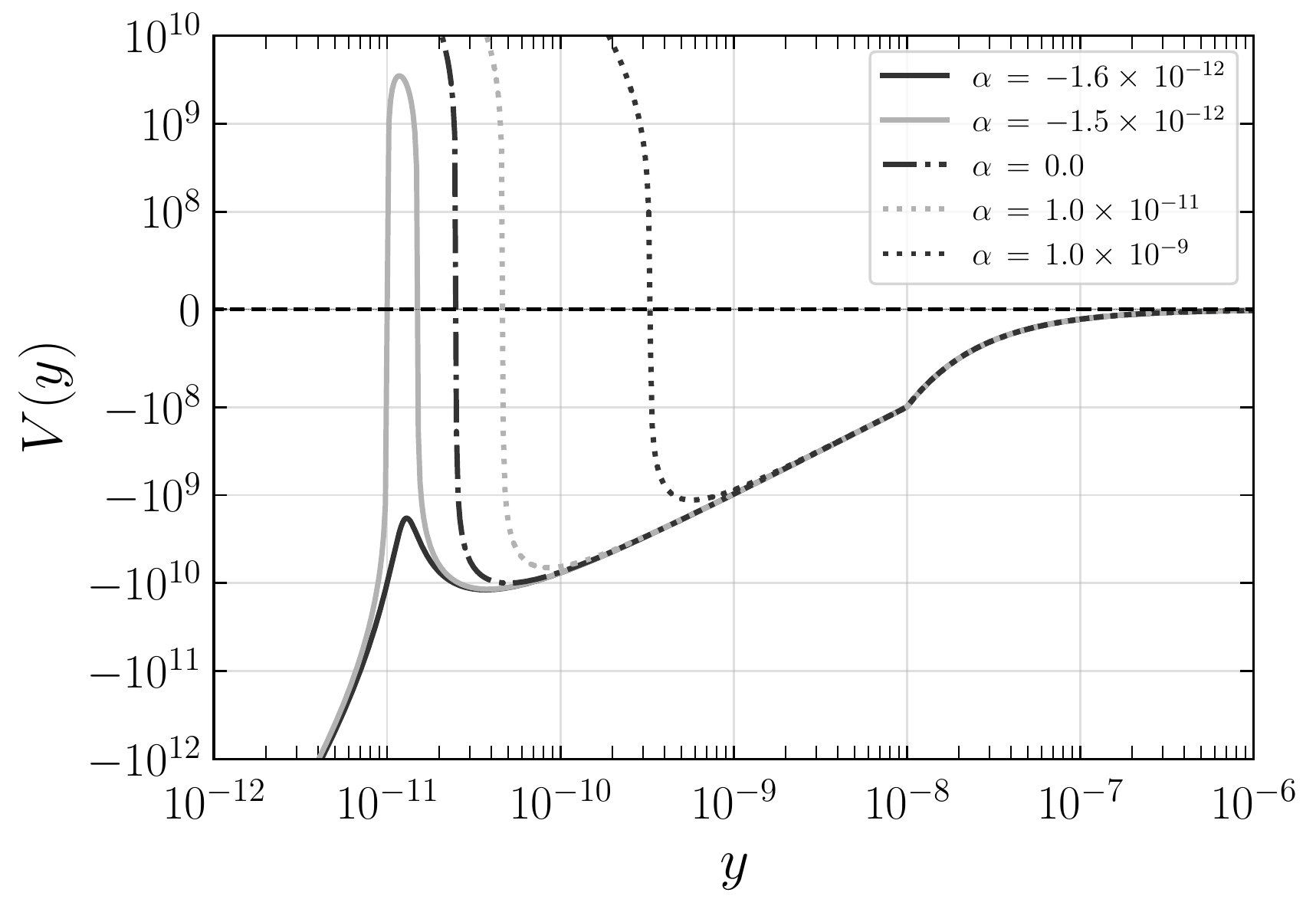}
 \caption{\label{fig:2}Behaviour of effective potential as a function of $y$ 
 for different values of $\alpha$. 
 Horizontal black dashed line denotes the effective energy of the fluid element of the shell with a 
 density profile specified by $\gamma\,=\,10^{-10}$ and with 
 initial conditions $\bar\delta_i\,=\,10^{-5}$ and $y_i\,=\,10^{-5}$ and 
 curves denote the effective potential of the fluid element for various values 
 of $\alpha$. 
 As we can see, for $\alpha = -1.6\times 10^{-12}$, 
 the potential does not have a region which is bounded from both sides 
 and hence the quantum pressure cannot stop the collapse of the shell. 
 For all other values of $\alpha$, shown in the figure, the shell will oscillate.
 }
\end{figure}
 \subsection{Effect of interactions}\label{subsec:interactions}
 We shall now try to understand the effect of interactions. 
 Due to the lack of analytical solution, we will approach the problem numerically. 
 For $a_s\,\neq\,0$, we can rewrite Eq. (\ref{eqn:eom}) in dimensionless form as
 \begin{equation}
   \f{\d^2 y}{\d\tau^2}\, =\, \f{\alpha\,(3\,\gamma\,-\,\gamma^2)}{y^4}\,+\, \f{2\,\gamma\, -\, \gamma^2}{4\,y^3}\, -\, \f{1}{y^2} \label{eqn:ywithas},
 \end{equation}
where, $a_s\, \equiv\, \alpha\, \bar{a}_s$ with $\bar{a}_s\, =\, \hbar^2/(G\,M^2\,m)$ and 
$\alpha$ can be greater than, equal to or less than zero which corresponds to 
repulsive, nil and attractive interaction respectively. 
In order to understand the effect of interactions, it is convenient to 
look at the form of the effective potential governing the evolution of the shell,
\begin{equation}
 V(y)\, =\, \f{\alpha\,(3\,\gamma\,-\,\gamma^2)}{3\,y^3}\,+\, \f{2\,\gamma\,-\gamma^2}{8\,y^3}\, -\, \f{1}{y}.
\end{equation}
\par
As one can see from Fig. \ref{fig:2}, the effect of non-zero interaction adds up with 
the quantum pressure when $\alpha >0$, whereas, it acts against quantum pressure for $\alpha < 0$. 
We can see that for $\alpha \lesssim -1.6\times10^{-12}$, the attractive force due to 
interaction and gravity is stronger than the repulsive force due to quantum pressure 
and hence the shell will collapse. 
Thus, the existence of a core will allow us to put a lower bound on $a_s$ for attractive interactions. 
For the parameters that we have considered, we will have a lower bound of $a_s\, \gtrsim\, -5.3\times 10^{-89}\,{\rm m}$. 
In other words, for attractive scattering length, $|a_s| << \bar{a}_s$ 
in order to sustain oscillations and hence form a core.
\begin{figure}
 \includegraphics[width = 12cm]{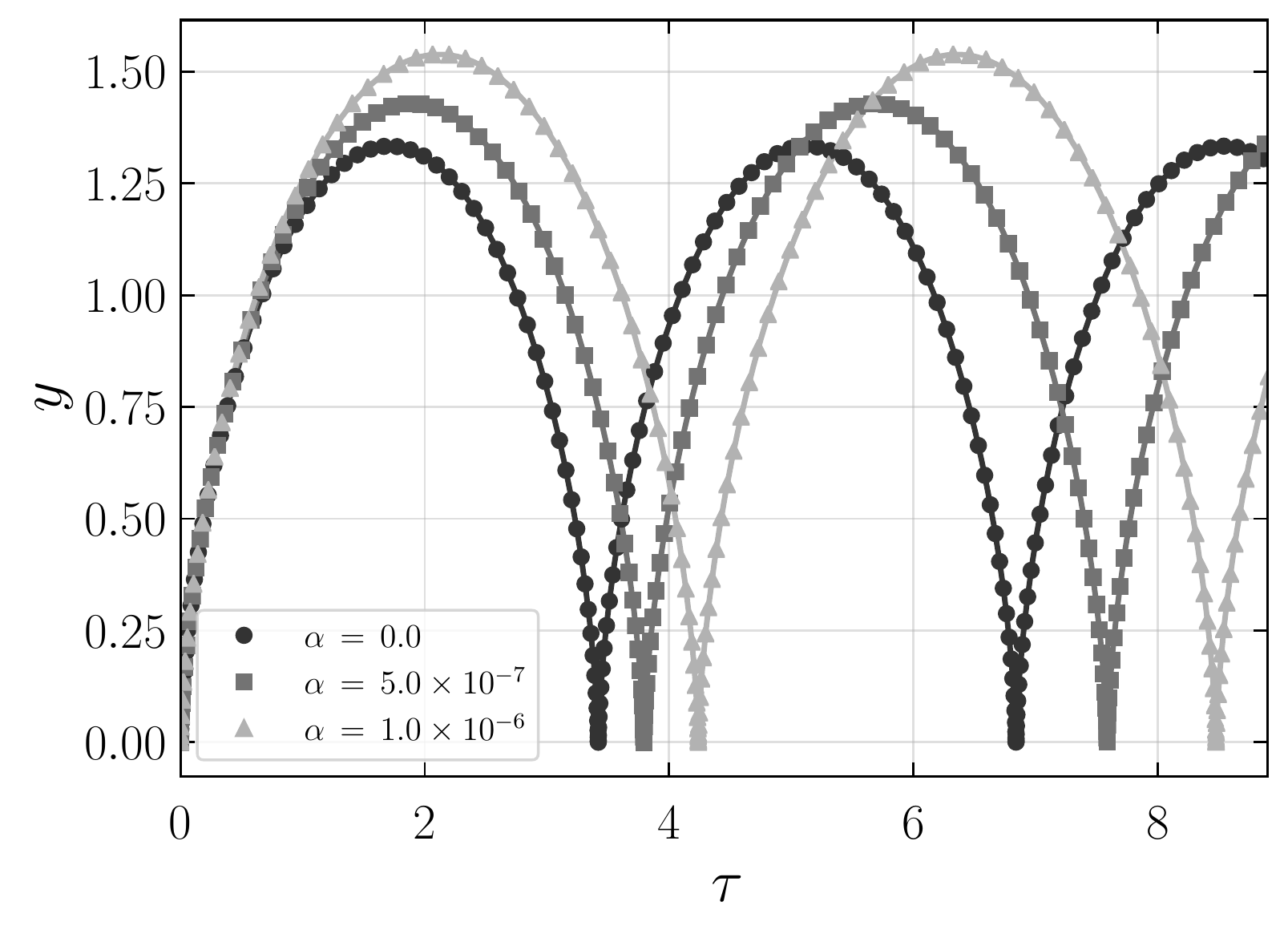}
 \caption{\label{fig:3}
 The evolution of the shell for 
 different values of $\alpha > 0$ are shown. 
 The effect of increasing $\alpha$ is a shift in the minimum of the potential to 
 larger values of $y$. 
 This would cause the fluid element to oscillate between larger 
 values of maximum and minimum and with a longer period. 
 The markers indicate the analytical expression Eq. (\ref{eqn:sol-Lt}).
 }
\end{figure}
\par
For $\alpha>0$, the effect of interactions is to push the minima of the potential to larger $y$. 
Hence, for larger $\alpha$, the shell would oscillate between larger maximum 
and minimum radius. 
In Fig. \ref{fig:3}, we have plotted the numerical evolution of the shell for various 
values of $\alpha$. 
The dots indicate the analytic expression Eq. (\ref{eqn:sol-Lt}). 
For the parameters that we use, the match is apparently good. 
This is because, in this toy model, the effect of interactions is only felt at small scales 
where as at large scales, the force is dominated by gravity. 
Though not evident from the plot, the minimum value of $y$ differs from 
the analytical value as the value of $\alpha$ is non-zero.
In particular, the analytical expression predicts a minimum radius of 
$y_{\rm min}\, =\, 2.5\times 10^{-11}$ where as the numerical simulations 
indicate a minimum radius of $1.5\times 10^{-11}$, $7.08\times 10^{-9}$ 
and $1.0\times 10^{-8}$ for $\alpha\, =\, -1.5\times 10^{-12},\, 5\times 10^{-7}$ 
and $10^{-6}$ respectively. 
Hence, one can conclude that, for the parameters that we have considered, the analytical expression derived in 
Section \ref{subsubsec:analytic} holds at large to medium scales 
for the case of interacting bosons. 
 \section{Virialization}\label{sec:virialization}
 The study of evolution of a single shell of radius $L$ of FDM containing a 
 mass $M$ shows that a sufficiently overdense shell would initially expand along 
 with the Hubble flow  and like in the case of CDM turn around and start contracting. 
 However, unlike CDM, instead of contracting to zero size, the shell may start expanding 
 again and repeat the process of expansion, turn around and contraction. 
 For the case of non-interacting bosons, the shell would always oscillate between 
 a minimum and maximum radius, as was shown in section \ref{subsubsec:analytic}. 
 In the case of interacting bosons, we had numerically shown that the shell would 
 oscillate only if the parameter $\alpha$ is above a certain value (see 
 section \ref{subsec:interactions}). 
 \par
 However, in reality, since different shells evolve at different rates, 
 as a shell is contracting, it will interact with inner shells which are 
 expanding again after their initial contraction. 
 When they interact, different shells will repel each other due to the quantum 
 pressure and repel or attract each other according to their force of interaction. 
 This would cause the density profile to depart from its initial power law shape (see, for instance, \cite{Kopp:2017hbb}). 
 Furthermore, at the time when radius of the shell reaches its minimum, the wave function 
 of the shell vanishes which might cause the Madelung transformation to break down (see, for instance, \cite{Uhlemann:2014npa}). 
 Thus, the shell would now have a more complicated dynamics which is not captured 
 by the equation of motion Eq. (\ref{eqn:eom}). 
 Nevertheless, since we know that the sphere of FDM would eventually virialize 
 to become a halo \cite{Schive:2014dra}, we can use the virial theorem to investigate beyond the validity 
 of Eq. (\ref{eqn:eom}). 
 For simplicity and because of the existence of an analytical solution, we will consider 
 the case of non-interacting bosons in this section.  
\par
The total energy of the system is given by
\begin{equation}
 E_{tot}\, =\, T\, +\, U_Q\, +\, U_I\, +\, U_G
\end{equation}
where, $T$ is the kinetic energy of the system, $U_Q$ is the energy stored in the 
system due to the quantum pressure, $U_I$ is the energy stored in the system 
due to the interaction and $U_G$ is the gravitational potential energy. 
When the system achieves virial equilibrium, the virial theorem states that (see, for instance, \cite{2011PhRvD..84d3531C})
\begin{equation}
 2\,T\, +\, 2\, U_Q\, +\, 3\,U_I\, +\, U_G\, =\, 0.
\end{equation}
For non-interacting bosons, the virial theorem hence implies that 
\begin{equation}
 T\, +\, U_Q\, =\, -U_G/2,
\end{equation}
which in turn implies that the total energy of the virialized halo 
is given by $E_{tot}\, =\, U_G/2$.
\par
At turn around, the energy of the system is dominated by the gravitational energy. 
This can be seen from the fact that the value of $\gamma$ in Eq. (\ref{eqn:y}) is small. 
Thus, at turn around, $E_{tot}\, \simeq\, U_G$. 
Using the fact that energy of the system is conserved and comparing the total energy 
at turn around and at virialization, one obtains, $L(t_{vir})\, =\, L(t_{ta})/2$, 
where we have used the expression for gravitational potential energy to be
\begin{equation}
 U_G\, =\, \f{\G\,M^2}{L(t)}\f{(3\,-\,\gamma)^2}{3\,(5\,-\,2\,\gamma)}.
\end{equation}
\par
Using the expression for radius of the shell at turn around 
and at virialization we will now compute the overdensity of the system at 
turn around and virialization in the full and the linear theory. 
At turn around, the overdensity in the full theory is computed using 
Eq. (\ref{eqn:overdensity}) and is given by,
\begin {equation}
 1\,+\,\bar\delta_{ta}\,=\, \f{9}{2}\,\f{(\pi\,-\,e\,\sin(\pi))^2}{(1\,-\,e\,\cos(\pi))^3}\, =\, \f{9}{2}\,\f{\pi^2}{(1+e)^3}.
\end {equation}
At virialization, the overdensity is given by 
\begin{equation}
  1\,+\,\bar\delta_{vir}\, =\, \f{9\, \G\, M}{2}\f{t_{vir}^2}{L(t_{vir})^3}.
\end{equation}
Using the expressions for the radius of the shell at turn around, and hence computing virial radius, $L_{vir}$, using $L_{vir}\,=\,L_{ta}/2$, one can compute the 
overdensity after virialization at $t_{vir}\,=\, t(2\,\pi)$ as
\begin{eqnarray}
 1\,+\,\bar\delta_{vir}\, &=&\, \f{9\,\G\,M}{2}\,\l[ \l(\f{A^3}{\G\,M}\r)^{1/2}\, (2\,\pi)\r]^2\,\times\,\f{8}{A^3\,(1\,+\, e)^3}\nonumber\\
 &=&\, 18\,\pi^2\,\f{8}{(1\,+\,e)^3}.
\end{eqnarray}
It can be verified that the averaged overdensity in the full theory matches with the CDM value in 
the $e\rightarrow 1$ limit. 
\par
Let us now compute the overdensity in the linear regime. 
As we discussed in section \ref{subsubsec:overdensity}, the small $\vartheta$ 
limit of $\bar\delta$ exists only in the limit $e\rightarrow 1$. 
In that limit the linear overdensity $\bar\delta\, \propto\, \vartheta^2$ as is 
shown in Eq. (\ref{eqn:lineardelta}). 
In order to express the linear overdensity in terms of time, 
we need to expand Eq. (\ref{eqn:sol-t}) in the $e\rightarrow 1$, 
$\vartheta\rightarrow 0$ limit. 
Expanding the expression for $t(\vartheta)$ in the $e\rightarrow1$, $\vartheta\rightarrow0$ limit, one obtains 
 \begin{equation}
 t\, \simeq\, \l(\f{m\,A^3}{k}\r)^{1/2}\f{e\,\theta^3}{6}.
\end{equation}
Using the above expression, one could write an expression for an overdensity at an initial time 
$t_i$ corresponding to $\vartheta_i$ as
\begin{equation}
 \bar\delta_i\, =\, \f{3\,\theta_i^2}{20}\, =\, \f{3}{20}\,\l[\f{6\,\pi}{e}\f{t_i}{t_{ta}}\r]^{2/3}.
\end{equation}
In an EdS universe, since $\delta\,\propto\, a$ in the linear regime, one could write an 
expression for $\delta(t)$ as 
\begin{eqnarray}
 \bar\delta\, &\propto&\, \bar\delta_i\,\f{a}{a_i}\,\nonumber\\
 &=&\, \f{3}{20}\, \l(\f{6\,\pi}{e}\r)^{2/3}\,\l(\f{t}{t_{ta}}\r)^{2/3}.
\end{eqnarray}
If we use the linear theory to compute the overdensity at turn around, one obtains,
\begin{equation}
 \bar\delta(t_{ta})\,\simeq\, \f{3}{20}\,\l(\f{6\,\pi}{e}\r)^{2/3}\, =\, \f{1.06}{e^{2/3}}.
\end{equation}
Upon using the linear theory to compute the overdensity after virialization, \ie at $t_{vir}\, =\, t(2\,\pi)$, 
we get
\begin{equation}
 \bar\delta(t_{vir})\, \simeq\, \f{3}{20}\,\l(\f{12\,\pi}{e}\r)^{2/3}\, =\, \f{1.69}{e^{2/3}}.
\end{equation}
The results obtained in this section have been summarized in Table \ref{table:1}. 
It shows that, in this model, when the linear averaged overdensity, 
reaches a critical value, $\bar\delta\,\simeq1.69/e^{2/3}\,\simeq\,1.69$, the 
overdense region would have virialized to form a halo. 
\begin{center}
\begin{table}
 \begin{tabular}{c|c|c}
  $t$ &
  Linear theory&
  Full theory\\
 & & \\
  \hline
& & \\
$t_{ta}$ & $\f{1.06}{e^{2/3}}$ &
$\f{9}{2}\,\f{\pi^2}{(1+e)^3}\,-\,1$\\ 
& & \\
%
$t_{vir}$ & $\f{1.69}{e^{2/3}}$ &
$18\,\pi^2\,\f{8}{(1\,+\,e)^3}\, -\, 1$\\
& & \\
%
  \end{tabular}  
  \caption{\label{table:1}The averaged overdensity, $\bar\delta$, in the linear 
  and the full theory at turn around and virialization. 
  It can be verified that the averaged overdensity matches with the CDM result 
  in the $e\rightarrow 1$ limit.}
 \end{table}
 \end{center}
 \section{Discussion}
 FDM is a compelling model for dark matter. 
 The quantum nature of FDM which gets manifested at kilo parsec scales is 
 capable of resolving the small scale issues that has been ailing CDM. 
 FDM halo can be described as a self-gravitating BEC and hence is governed by 
 the GPP equations (\ref{eqn:GPP}). 
 Numerical simulations \cite{Schive:2014dra} indicate that at large scales the structure formed in FDM is 
 similar to that in CDM. 
 High resolution simulations \cite{Schive:2014dra} show the existence of standing waves of dark matter 
 which evolves in to solitonic cores at the center of the halo. 
 As they accrete more matter, the solitonic core grows and are surrounded 
 by virialized halos with fine-scale, large-amplitude fringes. 
 The surrounding halos are supported against gravity by quantum and turbulent 
 pressure and hence fluctuates in density and velocity. 
 \par
 With the goal of gaining analytical insights in to the evolution of an FDM halo, 
we investigated the gravitational collapse of a spherical shell containing an overdense 
 region of FDM.
 We studied the system in its hydrodynamical form, \ie as a fluid with 
 density $\rho$ and velocity $\vu$ evolving under the effect of 
 opposing forces of Newtonian gravity and quantum pressure. 
 In an expanding universe, we computed the equation of motion 
 governing a spherical shell Eq. (\ref{eqn:R-general}).
 Assuming a spherically symmetric power law profile Eq. (\ref{eqn:pl-profile}) for the overdense region, 
 we derived an expression for the time evolution of the spherical shell Eqs. (\ref{eqn:sol-Lt})
 for the case of non-interacting bosons. 
 The correctness of the analytical solution was further established by 
 comparing it with numerical solution.
 Using the analytical expressions, we arrived at an expression for the averaged overdensity 
 enclosed by the spherical shell Eq. (\ref{eqn:overdensity}). 
 Further, we numerically evaluated the evolution of a shell in the presence of interaction 
 and compared it with the analytical expressions evaluated for the case of 
 non-interacting bosons. 
 \par
 We find that, as in the case of CDM, in the beginning, the spherical shell of FDM enclosing 
 the overdense region expands along with the Hubble flow and eventually turns around 
 and begin to collapse. 
 However, contrary to the case of CDM, due to the existence of quantum pressure, 
 the FDM spherical shell eventually gets repelled and starts expanding again. 
 A similar behaviour can be seen by looking at the expression of overdensity as well. 
 We can see from Eq. (\ref{eqn:overdensity}) that
 the overdensity remains finite and fluctuates with time. 
 The expression for overdensity also has the nice feature that it reproduces the CDM 
 result in the $\hbar/m \rightarrow 0$ limit. 
We further studied the initial linear evolution of the overdensity. 
It was found that, in this model, for an overdensity, a valid small $\vartheta$ limit exists only in the $e\rightarrow 1$ limit.
In the presence of interactions, the force due to interaction works along with the 
quantum pressure if the interaction is repulsive while acts against quantum pressure 
if it is attractive. 
We found that, for the parameters of interest, the spherical shell would oscillate  
in the case of attractive interaction, only if $|a_s| << \bar{a}_s$.
On the other hand, for repulsive interaction, the shell would oscillate with a larger 
maximum and minimum radius for larger values of $a_s$. 
We also found that, for the parameters that we have considered, which correspond to the 
numbers arrived for a dwarf spheroidal \cite{Schive:2014dra}, the analytical expression 
for the shell is a good approximation at large to medium scales. 
\par
As the shell contracts, it will interact with inner shells and the 
dynamics of the shell would be more complicated than the one 
captured by Eq. (\ref{eqn:sol-Lt}). 
In reality, as was shown in \cite{Schive:2014dra}, the spherical overdense 
region would eventually virialize to form a halo. 
Hence, the solutions discussed above, though captures 
some of the effects of various forces at play, will not be valid through out the evolution. 
However, we can investigate beyond the validity of the solutions Eq. (\ref{eqn:sol-Lt}), 
by making use of virial theorem. 
In section \ref{sec:virialization}, we used the virial theorem to compute the overdensity 
after virialization in the linear and in the full theory (see Table \ref{table:1}). 
In this model, as in the case of CDM \cite{paddylss1993}, we find the 
critical density at which the overdensity virializes to a halo to be $\bar\delta_c\,\simeq1.69/e^{2/3}\,\simeq\,1.69$. 
Finally, from the simulations performed for Fornax dwarf spheroidal, see Fig. \ref{fig:1}, our computations shows that
the virialized halo would have a radius of $L_{vir}\, \simeq\, L_{max}/2\, =\, .95\,\pc$. 
\par
 We shall conclude this article by discussing some of the subtleties involved in 
 the calculation and some interesting aspects that need to be further investigated.
 First of all, even though this study was motivated by the possibility of FDM being a
 viable dark matter candidate, the analytical calculations performed in this paper hold for any 
non-interacting BEC collapsing under the effect of gravity. 
Secondly, in the case of CDM, as the shell is contracting it will cross the shells which 
is expanding after their first in fall. 
In the case of FDM, however, when two shells come close to each other there will be 
repulsion due to the quantum pressure and hence the dynamics near shell crossing would 
be more involved than in CDM. 
Finally, in this work we have used the hydrodynamic description to model the system. 
It is not clear how well the hydrodynamic description captures the physics 
underlying the GPP equations (see, for instance, \cite{Uhlemann:2014npa, Kopp:2017hbb, Mocz:2018ium}). 
Hence, it would be interesting to investigate the regime close to the ``shell crossing'' in more detail.

 \section*{Acknowledgements}
  The author would like to thank T. Padmanabhan, Aseem Paranjape and L. Sriramkumar for extensive 
  discussions and comments. He would also like to thank the referee for useful 
  suggestions.
\bibliography{Refs}

\end{document}